# Deductive-reductive determination of the model of our observed Universe

Vladimír Skalský

Faculty of Materials Science and Technology of the Slovak Technical University, 917 24 Trnava, Slovakia, skalsky@mtf.stuba.sk

**Abstract.** According to the observations, in our expansive and isotropic relativistic Universe for the gravitational phenomena in a Newtonian approximation the Newtonian non-modified relations are valid. The Friedmann general equations of isotropic and homogeneous universe dynamics describe an infinite number of models of expansive and isotropic relativistic universe in the Newtonian approximation, but only in one of them the Newtonian non-modified relations are valid. These facts give – till now not considered – possibility for unambiguous deductive-reductive determination of the Friedmannian model, which describes our observed Universe.

## 1. Introduction

*The Einstein field equations* (Einstein 1915) we can express as:

$$G_{ik} \equiv R_{ik} - \frac{1}{2}g_{ik}R = \frac{8\pi G}{c^4}T_{ik}, \tag{1}$$

where $G_{ik}$ is the Einstein tensor, $R_{ik}$ is the Ricci tensor, $g_{ik}$ is the metric tensor, $R$ is the Ricci scalar and $T_{ik}$ is the energy-momentum tensor.

From the Einstein equations (1) we can derive *the Friedmann general equations of isotropic and homogeneous universe dynamics* (Friedmann 1922, 1924) in the form:

$$\dot{a}^2 = \frac{8\pi G a^2 \rho}{3} - kc^2 + \frac{\Lambda a^2 c^2}{3}, \tag{2a}$$

$$2a\ddot{a} + \dot{a}^2 = -\frac{8\pi G a^2 p}{c^2} - kc^2 + \Lambda a^2 c^2, \tag{2b}$$

where $a$ is the gauge factor, $\rho$ is the mass density, $k$ is the curvature index, $\Lambda$ is the cosmological member and $p$ is the pressure.

The equations (2a) and (2b) are mutually linked by *the Friedmannian general state equation*

$$p = K\varepsilon, \tag{2c}$$

where $K$ is the state equation constant and $\varepsilon$ is the energy density.

The Friedmann equations (Friedmann 1922, 1924) originated by Friedmann generalisation (Friedmann 1922, 1924) of *the Einstein first model of relativistic universe* (Einstein 1917) and *the de Sitter models of relativistic universe* (de Sitter 1917). From the mathematical-physical point of view, the Friedmannian equations (2a), (2b) and (2c) are applications of the Einstein equations (1) to the whole isotropic relativistic universe in the Newtonian (classical mechanics) approximation, under supplementary assumptions introduced into relativistic cosmology by Einstein (1917), de Sitter (1917) and Friedmann (1922, 1924), and generalised by Friedmann (1922, 1924).

The Friedmannian equations (2a), (2b) and (2c) have an infinite number of solutions, this means that they – without an introduction of any additional supplementary restrictive assumptions – describe infinite number of *the Friedmannian universe models*. Under the assumption that the mathematical-physical description of our Universe does not require introduction of any further supplementary conditions, this means that the Friedmannian equations (2a), (2b) and (2c) must immanently contain also a model which describes our observed expansive and isotropic relativistic Universe in the Newtonian approximation.

At present time a meaning prevails that the Friedmannian model describing our Universe can be determined only by observing its large-scale parameters. However, the chosen large-scale methods are of small precision and face colossal technical difficulties.

But there exists also another – till now not considered – the deductive-reductive possibility of determination of the Friedmannian model of universe describing our observed Universe, based on the following facts:

1. According to the observations, in our expansive and isotropic relativistic Universe for the gravitational phenomena in the Newtonian approximation, the Newtonian non-modified relations are valid.

2. Only in one of infinite number of the Friedmannian models of expansive isotropic and homogeneous

universe, described by the Friedmannian equations (2a), (2b) and (2c), the Newtonian non-modified relations are valid.

From these facts it results unambiguously that our observed Universe is described just by this only one – in present time unknown for us – Friedmannian model of universe with the Newtonian non-modified relations.

## 2. Deductive-reductive determination of the Friedmannian model of our Universe

In *the Einstein general theory of relativity* (Einstein 1915, 1916) and in *the Newton theory of gravitation (the classical mechanics)* (Newton 1687) for the local gravitational phenomena and for the whole of universe we use the same mathematical-physical apparatuses.

For the determination of the Friedmannian model of our observed Universe is significant mainly the fact that for some local gravitational phenomena expressed in the Newtonian approximation and for some global parameters of expansive and isotropic relativistic Universe expressed in the Newtonian approximation, the same Newtonian relations are valid.

In the Einstein general theory of relativity, the matter objects can move at the velocities $v$ of the interval $[0, c)$, where $c$ is *the boundary velocity of signal propagation*. From this fact it results that our observed expansive and isotropic relativistic Universe must be described by the relations with limit (boundary) relativistic matter (mass)-space-time values. This fact must be respected also by the model relations for the parameters of expansive and isotropic relativistic Universe expressed in the Newtonian approximation.

Therefore, in the deductive-reductive determination of the Friedmannian model describing our observed Universe in the Newtonian approximation, we shall proceed so that on the basis of the Newton theory of gravitation we derive the mass-space-time relations representing the relativistic limit relations in the Newtonian approximation, and using both of them and the Friedmannian equations (2a), (2b) and (2c), we determine the Friedmannian model of universe describing our observed Universe in the Newtonian approximation (Skalský 1997).

*The Newton law of gravitation* (Newton 1687) in present notation can be written as:

$$F = \frac{Gm_1 m_2}{r^2}, \qquad (3)$$

where $F$ is the gravitational force, $m_1$ and $m_2$ are the masses of spherically symmetrical bodies and $r$ is the distance of its centres.

From the relation (3) we can derive the relation for *the escape (second cosmic) velocity* $v_2$ in the gravitational field of spherically symmetrical body of mass $m$ at the distance $r$ from its centre:

$$v_2 = \sqrt{\frac{2Gm}{r}}. \qquad (4)$$

If in the relation (4) we substitute the escape velocity $v_2$ by the boundary velocity of signal propagation $c$, we receive the relation for *the Schwarzschild gravitational radius* (Schwarzschild 1916)

$$r_g = \frac{2Gm}{c^2}. \qquad (5)$$

If in the relation (5) we express the mass $m$ by the product of volume

$$V = \frac{4}{3}\pi r^3 \qquad (6)$$

and the mass density $\rho$, i.e.:

$$m = V\rho = \frac{4}{3}\pi r^3 \rho, \qquad (7)$$

we obtain the relation for the gravitational radius $r_g$ and the (critical) mass density $\rho$:

$$r_g = \sqrt{\frac{3c^2}{8\pi G\rho}}. \qquad (8)$$

The expansive isotropic and homogeneous universe model with the critical mass density $\rho$ is flat, therefore, for the gauge factor $a$ in it holds:

$$a = r_g. \qquad (9)$$

For the Hubble coefficient ("constant") $H$ and the cosmological time $t$ the well-known (non-modified Newtonian) relation (Hubble 1929) is valid:

$$H = \frac{1}{t}.\qquad(10)$$

The relations (5), (8), (9) and (10) determine the relations between parameters of the Newtonian model of the flat expansive isotropic and homogeneous universe (Skalský 1992, 1991):

$$a = ct = \frac{c}{H} = \frac{2Gm}{c^2} = \sqrt{\frac{3c^2}{8\pi G\rho}}.\qquad(11)$$

Using the Friedmannian equations (2a), (2b) and (2c) and the relations (11) we can determine the Friedmannian universe model, which describe our observed expansive and isotropic relativistic universe in the Newtonian approximation.

Individual Friedmannian models of expansive isotropic and homogeneous universe are unambiguously determined by the Friedmannian equations (2a), (2b) and (2c) with the values of curvature index $k = +1, 0, -1$, the values of cosmological member $\Lambda > 0, = 0, < 0$ and with the values of state equation constant $K > 0, = 0, < 0$.

The Newtonian model of the expansive isotropic and homogeneous universe with the relation between its parameters (11), is flat, therefore, in the Friedmannian description of it, by the Friedmannian relations (2a), (2b) and (2c), the curvature index $k$ in the relations (2a) and (2b) has the value $k = 0$ and the cosmological member $\Lambda$ in the relation (2a) and (2b) has the value $\Lambda = 0$.

The Friedmannian relations (2a) and (2b) with $k = 0$ and $\Lambda = 0$ and the relations (11) determine the state equation of the Friedmannian model of the flat expansive non-decelerative isotropic and homogeneous universe with the non-modified Newtonian relations (Skalský 1991):

$$p = -\frac{1}{3}\varepsilon.\qquad(12)$$

The gravitational force is determined by the sum of the energy density $\varepsilon$ and three-multiple of the pressure $p$. If in the relation (12) the value of energy density $-\varepsilon/3$ is shifted to the left side, it obtain the form:

$$\varepsilon + 3p = 0.\qquad(13)$$

It means that the relation (12) represents *the zero gravitational force state equation.*

Now we can state:

**The model of our observed expansive and isotropic relativistic Universe described in the Newtonian approximation, i.e.** *(the Friedmannian model of)* **the** *(flat)* **expansive non-decelerative** *(isotropic and homogeneous)* **universe** *(with the non-modified Newtonian relations)* **(ENU) is determined by the Friedmannian equations (2a), (2b) and (2c) with $k = 0$, $\Lambda = 0$ and $K = -1/3$.**

For better transparency, we express the relations (11) and (12) in all next possible relations and variants (Skalský 1997):

$$t = \frac{a}{c} = \frac{1}{H} = \frac{2Gm}{c^3} = \sqrt{\frac{3}{8\pi G\rho}},\qquad(14)$$

$$H = \frac{c}{a} = \frac{1}{t} = \frac{c^3}{2Gm} = \sqrt{\frac{8\pi G\rho}{3}},\qquad(15)$$

$$m = \frac{c^2 a}{2G} = \frac{c^3 t}{2G} = \frac{c^3}{2GH} = \sqrt{\frac{3c^6}{32\pi G^3 \rho}},\qquad(16)$$

$$\rho = \frac{3c^2}{8\pi G a^2} = \frac{3}{8\pi G t^2} = \frac{3H^2}{8\pi G} = \frac{3c^6}{32\pi G^3 m^2} = -\frac{3p}{c^2},\qquad(17)$$

$$p = -\frac{c^4}{8\pi G a^2} = -\frac{c^2}{8\pi G t^2} = -\frac{c^2 H^2}{8\pi G} = -\frac{c^8}{32\pi G^3 m^2} = -\frac{c^2\rho}{3} = -\frac{1}{3}\varepsilon.\qquad(18)$$

In the ENU all parameters are mutually linearly linked. Therefore, as a restricting condition for deriving the Friedmannian model of ENU not only the relations (5), (8), (9), or (10), need to be used, but as a restricting condition any of the relations (11) and (14)-(18) can be applied.

In the relations (17) and (18) we can ascertain that nor the state equation (12) has a privileged position in the ENU, and is merely one of the set of relations for linearly mutually linked parameters of the ENU.

The mass-space-time properties of our observed Universe you can see in the paper: *Matter-space-time properties of the Universe* (Skalský 1994) or in the paper: *The only non-contradictory model of universe* (Skalský 2000).

## 3. Conclusions

According to the observations – realised in the course of last five centuries – in our expansive and isotropic relativistic Universe for all gravitational phenomena in the Newtonian approximation, the Newtonian non-modified relations are valid.

The model of ENU is the only Friedmannian model of the expansive and isotropic relativistic universe in the Newtonian approximation, in which, for all gravitational phenomena in the Newtonian approximation, the Newtonian non-modified relations are valid.

From these facts it results unambiguously that our Universe in the Newtonian approximation has the properties of the Friedmannian model of ENU.

From the point of view of the Friedmannian cosmology this also means that the non-modified relativistic relations expressed in the Newtonian approximation (or the Newton non-modified gravitational law (3) and other Newtonian non-modified relations) are valid only in the Friedmannian model of the ENU, which is determined by the Friedmannian equations (2a), (2b) and (2c) with $k = 0$, $\Lambda = 0$ and $K = -1/3$, or by the definitive version of *the equations of universe dynamics* (Skalský 1993, 1997, 2000):

$$8\pi G a^2 \rho - 3c^2 = 0, \tag{19a}$$

$$8\pi G a^2 p + c^4 = 0, \tag{19b}$$

$$c^2 \rho + 3p = 0, \tag{19c}$$

or by their one-line variant:

$$\rho = \frac{3c^2}{8\pi G a^2} = -\frac{3p}{c^2}. \tag{20}$$

## References

De Sitter, W. 1917, *Proc. Acad. Wetensch. Amsterdam*, **19**, 1217.
Einstein, A. 1915, *Sitzber. Preuss. Akad. Wiss.* **48**, 844.
Einstein, A. 1916, *Ann. Phys.* **49**, 769.
Einstein, A. 1917, *Sitzber. Preuss. Akad. Wiss.* **1**, 142.
Friedmann, A. A. 1922, *Z. Phys.* **10**, 377.
Friedmann, A. A. 1924, *Z. Phys.* **21**, 326.
Hubble, E. P. 1929, *Proc. U.S. Nat. Acad. Sci.* **15**, 168.
Newton, I. 1960, *Philosophiae Naturalis Principia Mathematica*, Dawson, London
   (1687, first edition, Royal Society, London).
Schwarzschild, K. 1916, *Sitzber. Preuss. Akad. Wiss.* 189.
Skalský, V. 1991, *Astrophys. Space Sci.* **176**, 313. (Corrigendum: 1992, **187**, 163.)
Skalský, V. 1992, In: J. Dubnička (ed.), *Philosophy, Natural Sciences and Evolution* (Proceedings of an Interdisciplinary Symposium, Smolenice, December 12-14, 1988), Slovak Academy of Sciences, Bratislava, p. 83.
Skalský, V. 1993, *Astrophys. Space Sci.* **201**, 3. (Corrigendum: 1994, **219**, 303.)
Skalský, V. 1994, *Astrophys. Space Sci.* **219**, 275.
Skalský, V. 1997, *DYNAMICS OF THE UNIVERSE in the Consistent and Distinguished Relativistic, Classically-Mechanical and Quantum-Mechanical Analyses*, Slovak Technical University, Bratislava.
Skalský, V. 2000, astro-ph/… (To be published.)